\documentclass[12pt]{article}

\usepackage{adjustbox}
\usepackage{amsfonts}
\usepackage{amsmath}
\usepackage{amsthm}
\usepackage{amssymb}
\usepackage{authblk}
\usepackage{bm}
\usepackage{booktabs}  
\usepackage{caption}
\usepackage{diagbox}
\usepackage{enumerate}
\usepackage{float}
\usepackage{graphicx}
\usepackage{lineno}
\usepackage{listings}
\usepackage{mathtools}
\usepackage{mdframed}
\usepackage{multicol}
\usepackage{multirow}
\usepackage{natbib}
\usepackage{pdfpages}
\usepackage{rotating}
\usepackage{setspace}
\usepackage{subcaption}
\usepackage{threeparttable}  
\usepackage[normalem]{ulem}
\usepackage{url}
\usepackage{xcolor}

\newtheorem{example}{Example}
\newtheorem{theorem}{Theorem}
\newtheorem{definition}{Definition}
\newtheorem{proposition}{Proposition}

\graphicspath{{figures/}}

\begin{document}
	
\def\spacingset#1{\renewcommand{\baselinestretch}%
{#1}\small\normalsize} \spacingset{1}
	
\title{\bf Functional Bayesian Networks for Discovering Causality from Multivariate Functional Data}
\author{Fangting Zhou$^{1,2}$, Kejun He$^{2,\ast}$, Kunbo Wang$^{3}$, Yanxun Xu$^{3}$, and Yang Ni$^{1,\ast}$ \\
$^{1}$Department of Statistics, Texas A\&M University, College Station, Texas, U.S.A. \\
$^{2}$Institute of Statistics and Big Data, Renmin University of China, Beijing, China \\
$^{3}$Department of Applied Mathematics and Statistics, Johns Hopkins University, Baltimore, Maryland, U.S.A. \\
Email: kejunhe@ruc.edu.cn, yni@stat.tamu.edu}
\date{}

\maketitle

\begin{abstract}
Multivariate functional data arise in a wide range of applications. One fundamental task is to understand the causal relationships among these functional objects of interest, which has not yet been fully explored. In this article, we develop a novel Bayesian network model for multivariate functional data where the conditional independence and causal structure are both encoded by a directed acyclic graph. Specifically, we allow the functional objects to deviate from Gaussian process, which is adopted by most existing functional data analysis models. The more reasonable non-Gaussian assumption is the key for unique causal structure identification even when the functions are measured with noises. A fully Bayesian framework is designed to infer the functional Bayesian network model with natural uncertainty quantification through posterior summaries. Simulation studies and real data examples are used to demonstrate the practical utility of the proposed model.
\end{abstract}


\section{Introduction} \label{sec::introduction}

This article develops a novel functional Bayesian network for modeling directed conditional independence and causal relationships of multivariate functional data, which arise in a wide range of applications. For example, learning brain effective connectivity networks from electroencephalogram (EEG) records is crucial for understanding brain activities and neuron responses. Another example is longitudinal medical studies where multiple clinical variables are recorded at possibly distinct time points across variables and/or patients. Knowing causal dependence of these clinical variables may help physicians decide the right interventions. Functional data can also go beyond those defined on time domain e.g., spatial domain (environmental data, spatially-resolved genomics, etc).

Joint analysis of multiple functional objects has attracted great attention in recent years with focuses mainly on reducing dimensionality and capturing functional dependence. For instance, \cite{kowal2017bayesian} and \cite{kowal2019integer} proposed to model time-ordered functional data through a time-varying parameterization for functional time series. Using basis transformation strategies, \cite{zhang2016functional} built an autoregressive model for spatially correlated functional data, while \cite{lee2018bayesian} modeled functional data in serial correlation semiparametrically. \cite{chiou2014linear} developed a linear manifold model characterizing the functional dependence between multiple random processes.

\paragraph{Functional Graphical Models} In a similar but conceptually different manner, functional graphical models have been recently proposed to model conditional independence of multivariate functional data. Graphical models gives rise to compact probabilistic representation of high-dimensional data through the graph-encoded conditional independence constraints. One key challenge is that the graph is typically unknown and must be inferred from data. While graphical models have been extensively studied for vector- and matrix-variate data \citep{yuan2007model, wang2009bayesian, leng2012sparse, ni2017sparse}, only recently have there been several developments for the functional data. \cite{zhu2016bayesian} extended Markov and hyper Markov laws of decomposable undirected graphs for random vectors to those for random functions. \cite{qiao2019functional} adopted the group lasso penalty on the precision matrix of coefficients extracted from the basis expansion of functions. \cite{zapata2022partial} introduced the idea of partial separability to reduce the computational cost of \cite{qiao2019functional}. \cite{qiao2020doubly} further extended \cite{qiao2019functional} and proposed to characterize the time-varying conditional independence of random functions through smoothing techniques. To relax the Gaussian process assumption of the aforementioned methods, \cite{li2018nonparametric}, \cite{solea2022copula}, and \cite{lee2022nonparametric} proposed models based on additive conditional independence and copula Gaussian models.

\begin{figure}[h]
\centering
\begin{subfigure}[h]{0.25\textwidth}
\includegraphics[width=\textwidth]{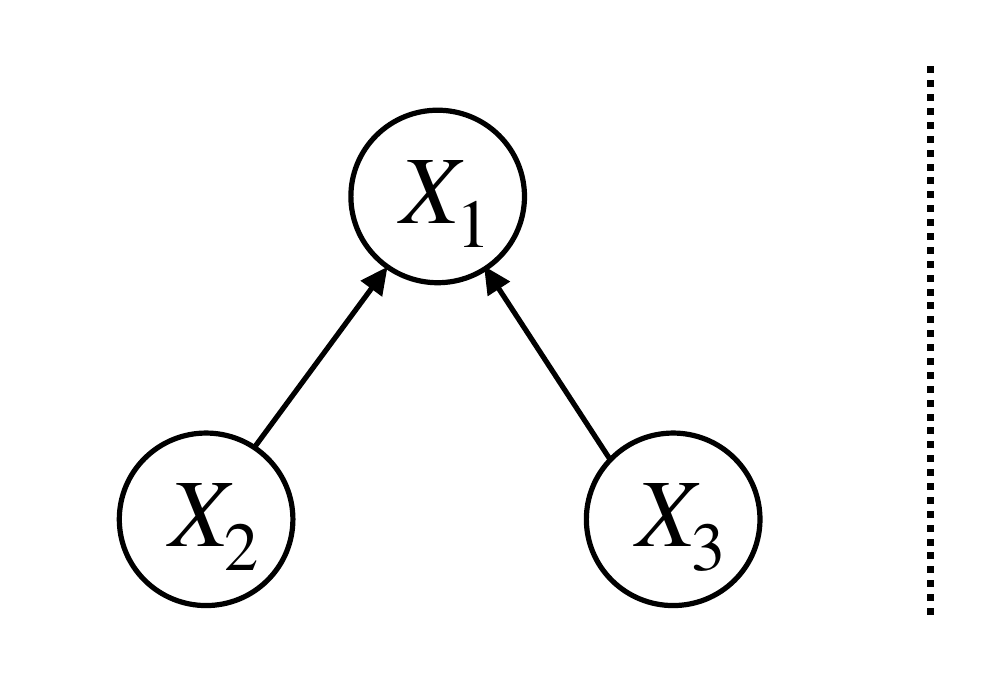}
\caption{}
\label{ex1}
\end{subfigure}
\begin{subfigure}[h]{0.72\textwidth}
\includegraphics[width=\textwidth]{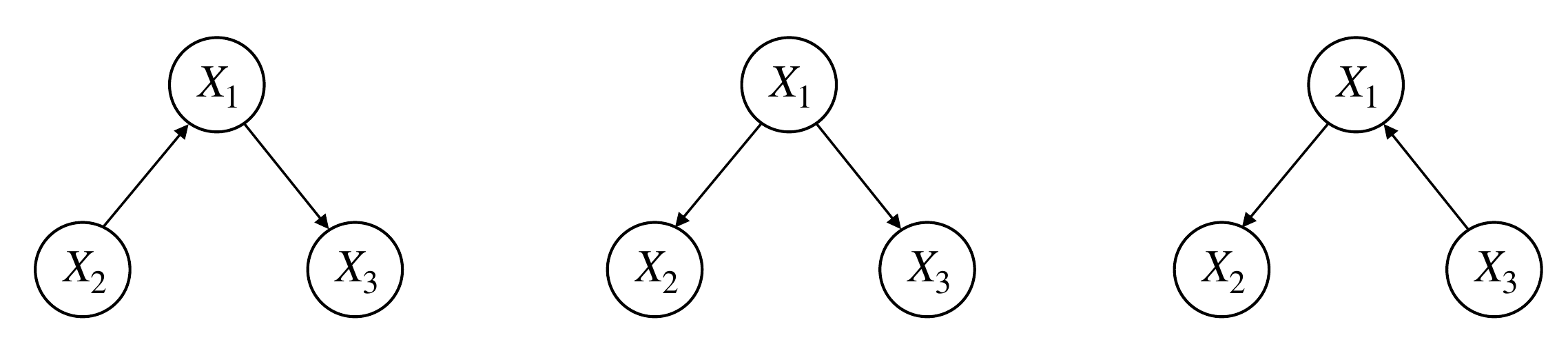}
\caption{}
\label{ex2}
\end{subfigure}
\caption{Two Markov equivalence classes. (a) $X_2 \perp X_3$. (b) $X_2 \perp X_3 | X_1$.}
\end{figure}

Despite these exciting developments of functional undirected graphical models, the work on functional \textit{directed} graphical models is sparse. Generally, undirected graphs admit a different set of conditional independence constraints from directed graphs. For example, the directed graph in Figure \ref{ex1} implies $X_2 \perp X_3$ but $X_2 \not\perp X_3 | X_1$, yet there exists no undirected counterpart that admits the same set of conditional (in)dependence assertions. More importantly, causal discovery (i.e., generation of plausible causal hypotheses) is only possible with directed graphs given additional causal assumptions \citep{pearl2000causality}. To the best of our knowledge, the functional structural equation model recently proposed by \cite{lee2022functional} is the only work that infers directional relationships from multivariate functional data. However, as will become evident in Section \ref{sec::fbn} and \ref{sec::inference}, our model differs from theirs in several significant aspects.

\paragraph{Causal Discovery} As hinted earlier, one of the two important problems we intend to address in this work is discovering causality from functional observations. Causal discovery is one of the first steps to investigate the physical mechanism that governs the operation and dynamics of an unknown system. Given the learned causal knowledge, subsequent causal inference (e.g., deriving the interventional and counterfactual distributions) can be conducted under the celebrated do-calculus framework \citep{pearl2000causality}. Therefore, inferring causal relationships potentially has more significant scientific impacts than learning associations since it may help answer fundamental questions about the nature. Bayesian networks paired with causal assumptions are among the most popular approaches in identifying unknown causal structure represented by a directed acyclic graph (DAG). One pressing obstacle of using Bayesian networks to discover causality from purely observational data is that in general, only Markov equivalence classes (MEC) can be learned based on conditional independence constraints alone. Causal interpretations of members in the same MEC can be drastically different, and, generally, only bounds on causal effects can be calculated \citep{maathuis2009estimating}. For example, the three DAGs in Figure \ref{ex2} constitute an MEC with the only conditional independence $X_2 \perp X_3 | X_1$, but the causal directions are completely reversed in the last graph compared to the first one.

Since 2006, numerous researchers, however, have found that causal discovery (unique causal structure identification) is indeed possible with additional distributional assumptions on the data generating process, at least for finite-dimensional data. Examples include but are not limited to linear non-Gaussian models (LiNGAM, \citealt{shimizu2006linear}), non-linear additive noise models \citep{hoyer2008nonlinear}, and linear Gaussian models with equal error variances \citep{peters2014identifiability}. See more related methods in a recent book of \cite{peters2017elements}. Although remarkable progresses have been made in the causal discovery area for traditional finite-dimensional data, what remains lacking is method capable of discovering causality from general, purely observational, multivariate functional data. We remark that given a known causal graph, there are existing approaches that can be used to infer causal effects. For example, \cite{lindquist2012functional} developed a causal mediation analysis framework where the treatment and outcome are scalars and the mediator is a univariate random function. Our scope is substantially different from this line of works in that we do not assume the causal graph to be known; in fact, learning the causal graph structure is precisely the focus of this paper.

\paragraph{Proposed Functional Bayesian Networks} We propose a novel functional Bayesian network model for multivariate functional data for which the conditional independence and causal relationships are represented by a DAG. As one would expect, the proposed functional Bayesian network factorizes over the DAG and respects all directed Markov properties (i.e., conditional independence constraints) encoded in the DAG via the notion of d-separation. Then for ease of exposition, we reformulate the proposed Bayesian network constructed in the functional space to an equivalent Bayesian network defined on the space of basis coefficients via basis expansion. Because in practice, functional data are almost always observed with noises, two essential ingredients are built in the proposed Bayesian networks to capture the functional dependence and to learn the causal structure. First, we capture the within-function dependence through a set of orthonormal basis functions chosen in a data-driven way. The resulting basis functions are interpretable and computationally efficient. Second, we encode the unknown causal structure by a structural equation model on the basis coefficients. Due to the equivalence of probability measures on the functional space and the space of basis coefficients, the conditional independence and causal relationships naturally transform back to the original random functions. To allow for unique DAG identification, we move away from the Gaussian process assumption often adopted by the existing functional graphical models and instead assume our random functions are generated from a discrete scale mixture of Gaussian distributions. We theoretically prove and empirically verify that the unique DAG identification is indeed possible even when the functions are observed with noises.

To conduct inference and uncertainty quantification from a finite amount of data, the proposed model is based on a Bayesian hierarchical formulation with carefully chosen prior distributions. Posterior inference is carried out through Markov chain Monte Carlo (MCMC). We perform simulation studies to demonstrate the capability of the proposed  model in recovering causal structure and key parameters of interest. A real data analysis with brain EEG records illustrates the applicability of the proposed framework in real world. We also apply the proposed model to a COVID-19 multivariate longitudinal dataset (shown in Section D of the Supplementary Material).

The rest of the paper is structured as follows. We provide an overview of Bayesian networks in Section \ref{sec::overview}. The proposed functional Bayesian network is introduced in Section \ref{sec::fbn}, which includes elaborations of the functional linear non-Gaussian model (Section \ref{sec::FLiNG}) and the causal identifiability theory (Section \ref{sec:ci}). Section \ref{sec::inference} is devoted to Bayesian inference of the proposed model. We provide simulation studies and applications in Sections \ref{sec::experiment} and \ref{sec::eeg}, respectively. The main contributions of this paper are summarized in Section \ref{sec::discussion} with some concluding remarks.
 
\section{Overview of Bayesian Networks} \label{sec::overview}

Throughout the paper, vectors and matrices are boldfaced whereas scalars and sets are not.

\paragraph{DAGs and Bayesian Networks} Let $\bm{X} = (X_1, \ldots, X_p)^T \in \mathcal{X}_1 \times \cdots \times \mathcal{X}_p$ denote a $p$-dimensional random vector. Denote $[m]: = \{1, \ldots, m\}$ for any integer $m \geq 1$. Let $\bm{X}_S = (X_j)_{j\in S}$ be a subvector of $\bm{X}$ with $S\subseteq [p]$. A DAG $G = (V, E)$ consists of a set of nodes $V = [p]$ and a set of directed edges represented by a binary adjacency matrix $\bm{E} = (E_{j\ell})$ where $E_{j\ell} = 1$ if and only if $\ell\rightarrow j$ for $\ell \neq j\in V$. DAGs do not allow directed cycles $j_0 \to j_1 \to \cdots \to j_k = j_0$. Each node $j \in V$ represents a random variable $X_j \in \mathcal{X}_j$; we may use $j$ and $X_j$ interchangeably when no ambiguity arises. Each directed edge $\ell \to j$ and the lack thereof represent conditional dependence and independence of $X_\ell$ and $X_j$, respectively. Note that although $X_j$ is often a scalar but it does not need to be. In fact, $X_j$ is a random function or an infinite dimensional random vector in this article. Denote $pa_G(j) = \{\ell\in V: \ell \to j\}$ the set of parents of $j$ in graph $G$. A Bayesian network (BN) $\mathcal{B} = (G, P)$ on $\bm{X}$ is a probability model where the joint probability distribution $P$ of $\bm{X}$ factorizes with respect to $G$ in the following manner,
\begin{equation}\label{eq:bnf}
P(\bm{X}) = \prod_{j = 1}^p P_j(X_j | \bm{X}_{pa_G(j)}),
\end{equation}
where $P_j$ is the conditional distribution of $X_j$ given $\bm{X}_{pa_G(j)}$ under $P$. Let $de_G(j) = \{\ell \in V: j \to \cdots \to \ell\}$ denote the descendants of $j$ in $G$ and let $nd_G(j) = V \backslash de_G(j) \backslash \{j\}$ denote the non-descendants of $j$. The BN factorization \eqref{eq:bnf} directly implies the local directed Markov property -- any variable is conditionally independent of its non-descendants given its parents, $X_j \perp \bm{X}_{nd_G(j)/pa_G(j)} | \bm{X}_{pa_G(j)}, \forall j \in [p]$. In fact, the reverse is also true: if a distribution $P$ respects the local Markov property according to a DAG $G$, then $P$ must factorize over $G$ as in \eqref{eq:bnf}. In summary, BN factorization and local Markov property are equivalent. We may omit the subscript $G$ of $pa_G(j)$ and $nd_G(j)$ and simply write $pa(j)$ and $nd(j)$ instead when $G$ is clear from the context.

\paragraph{Causal DAGs and Causal Bayesian Networks} A causal DAG $G$ is a DAG except that the directed edges are now interpreted causally, i.e., we say $X_\ell$ is a direct cause (with respect to $V$) of $X_j$ and $X_j$ is a direct effect of $X_\ell$ if $\ell \to j$. For simplicity, we will overload $nd(j)$ and $pa(j)$ to denote the noneffects and directed causes of $j$ in a causal DAG. To define a causal BN, we begin by asserting the local causal Markov assumption \citep{spirtes2000causation,pearl2000causality} -- given a causal DAG $G$, a variable is conditionally independent of its noneffects given its direct causes. By noting the correspondence between noneffects and non-descendants, and between direct causes and parents in DAGs and causal DAGs, the local causal Markov assumption simply states that the distribution $P$ of $\bm{X}$ respects the local Markov property of the causal DAG $G$, which in turn implies that $P$ must also factorize over $G$ (recall the equivalence between BN factorization and local Markov property). Therefore, a causal BN $\mathcal{B} = (G, P)$ is a probability model where $P$ factorizes with respect to a causal DAG $G$ in the same way as in \eqref{eq:bnf}.

\paragraph{Structural Equation Representation of Bayesian Networks} A BN is often represented by a structural equation model (SEM),
\begin{equation*}
X_j = f_j(\bm{X}, \epsilon_j), ~ \forall j \in [p],
\end{equation*}
where the transformation $f_j$ depends on $\bm{X}$ only through its parents/direct causes $\bm{X}_{pa(j)}$, and the exogenous variables $\bm{\epsilon}=(\epsilon_1,\dots,\epsilon_p)^T\sim P_\epsilon$ are assumed to be mutually independent. Denote the set of transformation functions as $F = \{f_1, \ldots, f_p\}$. Since $F$ and $P_\epsilon$ induce the joint distribution $P$ of $\bm{X}$ and it is not difficult to show that the induced distribution $P$ factorizes over $G$, with a slight abuse of notation, we can rewrite the BN as $\mathcal{B} = (G, F, P_\epsilon)$.
 
\section{Functional Bayesian Networks} \label{sec::fbn}

\subsection{General Framework} \label{sec:gf}

Now we introduce the construction of BNs for multivariate functional data. Denote the space of square integrable functions on domain $\mathcal{D}$ with respect to measure $\mu$ as $L^2(\mathcal{D}) = \{h: \int_\mathcal{D} h^2(\omega) d\mu(\omega) < \infty\}$. We focus on a compact $\mathcal{D} \subset \mathbb{R}$ (in fact, without loss of generality, $\mathcal{D} = [0, 1]$) and the Lebesgue measure $\mu$ for simplicity. Let $\bm{Y} = (Y_1, \ldots, Y_p)^T \in L^2(\mathcal{D}_1)\times\dots\times L^2(\mathcal{D}_p)$ be a collection of $p$ random functions. Denote $\mathcal{H} = \bigcup_{j = 1}^p \{(\omega, j): \omega \in \mathcal{D}_j\}$ the joint domain of $\bm{Y}$ and $(L^2(\mathcal{H}), \mathcal{B}(L^2(\mathcal{H})), P)$ its probability space. Similarly, for any subset $A \subset [p]$, denote the joint domain $\mathcal{H}_A = \bigcup_{j \in A} \{(\omega, j): \omega \in \mathcal{D}_j\}$ and $\mathcal{B}(L^2(\mathcal{H}_A))$ the Borel $\sigma$-algebra on $L^2(\mathcal{H}_A)$. Let $A, B, C$ be disjoint subsets of $[p]$. Following \cite{zhu2016bayesian}, we say $\bm{Y}_A$ is conditionally independent of $\bm{Y}_B$ given $\bm{Y}_C$ under $P$, if for any measurable set $D_A \subset L^2(\mathcal{H}_A)$, $P(\bm{Y}_A \in D_A | \bm{Y}_B, \bm{Y}_C)$ is $\mathcal{B}(L^2(\mathcal{H}_C))$ measurable and $P(\bm{Y}_A \in D_A | \bm{Y}_B, \bm{Y}_C) = P(\bm{Y}_A \in D_A | \bm{Y}_C)$. We introduce a DAG $G = (V, E)$ where each node $j \in V$ represents a random function $Y_j$. To begin with, we give the formal definition of a functional Bayesian network.

\begin{definition}[Functional Bayesian Networks]
We say $\mathcal{B} = (G, P)$ is a functional Bayesian network for a set of random functions $\bm{Y}$ if $P$ factorizes with respect to DAG $G$, 
\begin{align*}
P(Y_1 \in D_1,\dots,Y_p\in D_p) = \prod_{j = 1}^p P_j(Y_j \in D_j | \bm{Y}_{pa(j)} \in D_{pa(j)}),
\end{align*}
for any measurable sets $D_j \subset L^2(\mathcal{D}_j), \forall j \in [p]$, where $P_j$ is the conditional probability measure of $Y_j$ given $\bm{Y}_{pa(j)}$ under $P$.
\end{definition}

Just like the ordinary finite-dimensional BN, the functional BN factorization implies the local Markov property and vice versa. 

\begin{definition}[Functional Local Directed Markov Property]
A probability measure $P$ of $\bm{Y}$ satisfies the local directed Markov property with respect to $G$ if $Y_j \perp \bm{Y}_{nd(j)/pa(j)} | \bm{Y}_{pa(j)}$, i.e., $P(Y_j \in D_j | \bm{Y}_{nd(j)/pa(j)}, \bm{Y}_{pa(j)})$ is $\mathcal{B}(L^2(\mathcal{H}_{pa(j)}))$ measurable and $P(Y_j \in D_j | \bm{Y}_{nd(j)/pa(j)}, \bm{Y}_{pa(j)}) = P(Y_j \in D_j | \bm{Y}_{pa(j)})$ for any $D_j\subset L^2(\mathcal{D}_j)$.
\end{definition}

\begin{proposition}
Functional Bayesian network factorization is equivalent to functional local directed Markov property.
\end{proposition}

Proof is trivial. For modeling convenience, we use orthonormal basis expansion of random functions to (equivalently) redefine the functional BN in the space of basis coefficients. Let $\{\phi_{jk}\}_{k = 1}^\infty$ be a sequence of orthonormal basis functions of $L^2(\mathcal{D}_j)$ and expand $Y_j = \sum_{k = 1}^\infty Z_{jk} \phi_{jk}$, where $Z_{jk} = \int_{\mathcal{D}_j} Y_j(\omega) \phi_{jk}(\omega) d\omega$. The resulting coefficient sequence $\bm{Z}_j = (Z_{jk})_{k = 1, \ldots, \infty}$ lies in the space of square summable sequences $\ell_j^2 = \{h_j: \sum_{k = 1}^\infty h_{jk}^2 < \infty\}$. The within-function and the between-function covariance can then be expressed in terms of the covariance of the coefficient sequences,
\begin{align*}
\text{cov}(Y_j(\omega_j), Y_\ell(\omega_\ell)) = \sum_{k = 1}^\infty \sum_{h = 1}^\infty \phi_{jk}(\omega_j) \phi_{\ell h}(\omega_\ell) \text{cov}(Z_{jk}, Z_{\ell h}), ~ \forall \omega_j \in \mathcal{D}_j, \omega_\ell \in \mathcal{D}_\ell, ~ \forall j, \ell \in [p].
\end{align*}
Because $L^2(\mathcal{D}_j)$ and $\ell_j^2$ are isometrically isomorphic for each $j$, for any disjoint subsets $A, B, C \subset [p]$, $\bm{Y}_A \perp \bm{Y}_B | \bm{Y}_C$ if and only if $\bm{Z}_A \perp \bm{Z}_B | \bm{Z}_C$ where $\bm{Z} = (\bm{Z}_1, \ldots, \bm{Z}_p)^T$. Hence, if $\bm{Y}$ follows the proposed BN model $\mathcal{B} = (G, P)$, then the coefficient sequences $\bm{Z}$ follows $\mathcal{B}_Z = (G, P_Z)$ for some probability measure $P_Z$ of $\bm{Z}$, and vice versa. Each node of the DAG $G$ either represents a random function $Y_j$ or, equivalently, its corresponding coefficient sequence $\bm{Z}_j$. Moreover, the joint probability measure $P$ of $\bm{Y}$ factorizes with respect to $G$ if and only if the joint probability measure $P_Z$ of $\bm{Z}$ factorizes with respect to $G$. 

\begin{proposition}
Suppose $\bm{Y}\sim P$ and let $\bm{Z}$ be the corresponding coefficient sequences from orthonormal basis expansion.
Then 
\begin{align*}
P(Y_1 \in D_1,\dots,Y_p\in D_p) = \prod_{j = 1}^p P_j(Y_j \in D_j | \bm{Y}_{pa(j)} \in D_{pa(j)}),
\end{align*}
for any measurable sets $D_j \subset L^2(\mathcal{D}_j), \forall j \in [p]$ if and only if
\begin{align*}
P_Z(\bm{Z}_1 \in D_1',\dots,\bm{Z}_p\in D_p') = \prod_{j = 1}^p P_{Z j}(\bm{Z}_j \in D_j' | \bm{Z}_{pa(j)} \in D_{pa(j)}'),
\end{align*}
for any measurable sets $D_j' \subset \ell_j^2, \forall j \in [p]$.
\end{proposition}

The proof directly follows the preceding paragraph. Again, just like the ordinary finite-dimensional BN, if one makes the causal Markov assumption, the DAG $G$ in the proposed functional BN can be interpreted causally. Hereafter, by default, we always make the causal Markov assumption (hence $G$ is a causal DAG, the edge strength is interpreted as direct causal effect, etc) but all the results are simply reduced to those of a directed conditional independence model when the causal Markov assumption is dropped.
 
\subsection{Functional Linear Non-Gaussian Bayesian Networks} \label{sec::FLiNG}

Section \ref{sec:gf} introduces a general framework for modeling directed conditional independence and causal relationships for multivariate functional data. In this subsection, we discuss in detail one specific case of the proposed general framework, namely the \underline{F}unctional \underline{Li}near \underline{N}on-\underline{G}aussian (FLiNG) BNs. Specifically, the FLiNG-BN assumes $\bm{Z}$ follows a linear SEM,
\begin{align} \label{eq1}
\bm{Z}_j = \sum_{\ell = 1}^p \bm{B}_{j\ell} \bm{Z}_\ell + \bm{\epsilon}_j, ~ \forall j \in [p],
\end{align}
where $\bm{\epsilon}_j$ is an infinite-dimensional exogenous vector, $\bm{B}_{j\ell} = (B_{j\ell}(k_j, k_\ell))_{k_j=1,k_\ell=1}^{\infty,\infty}$ is an infinite-dimensional direct causal effect matrix from $\bm{Z}_\ell$ to $\bm{Z}_j$, and $\ell\to j$ is present in $G$ (i.e., $\bm{Z}_\ell$ is a direct cause of $\bm{Z}_j$) if there exist $k_\ell$ and $k_j$ such that $B_{j\ell}(k_j, k_\ell) \neq 0$. Neither causal effects nor the causal graph is assumed to be known; therefore the main goal of this article is precisely to infer them from observational data. Because $L^2(\mathcal{D}_j)$ and $\ell_j^2$ are isometrically isomorphic for all $j \in [p]$, the casual relationships of $\bm{Z}$ encoded in DAG $G$ directly transfer to the the casual relationships of $\bm{Y}$, i.e., $\bm{Z}_\ell$ is a direct cause of $\bm{Z}_j$ if and if only if $Y_\ell$ is a direct cause of $Y_j$.

In practice, the random functions $\bm{Y}$ can only be measured on finite grids with random noises. In other words, we do not observe realizations of $\bm{Y}$ but instead we observe realizations of $\bm{W}=(\bm{W}_1,\dots,\bm{W}_p)^T$ where $\bm{W}_{j} = (W_{j}(1), \ldots, W_{j}(m_j))$, which is the set of measurements of $Y_{j}$ on a finite grid $D_{j} = \{\omega_{j}(1), \ldots, \omega_{j}(m_j)\} \subset \mathcal{D}_j$ with independent white noises $e_{j}(m) \sim N(0, \sigma_j)$, $\forall m \in [m_j]$, 
\begin{align} \label{eq2}
W_j(m) = Y_{j}(\omega_{j}(m)) + e_{j}(m).
\end{align}
Note that $D_{j}$ can be different across $j$ (also across realizations).

One seemingly inconsequential element of the FLiNG-BN but turning out to be crucial for discovering causality is the specification of the probability distribution of the exogenous variables $\bm{\epsilon}_j= (\epsilon_{jk})_{k=1}^\infty$ in \eqref{eq1}.
A tempting choice may be Gaussian but it is the non-Gaussianity of $\bm{\epsilon}_j$'s that allows causal identification as we will show in Section \ref{sec:ci}. Specifically, we assume $\epsilon_{jk}$ to follow a finite scale mixture of Gaussian distributions, $\epsilon_{jk} \sim \sum_{m = 1}^{M_{jk}} \pi_{jkm} N(0, \tau_{jkm})$,
where $M_{jk}$ is the number of mixture components. The non-Gaussian exogenous variables lead to non-Gaussian coefficient sequences $\bm{Z}$, which in turn lead to non-Gaussian-process distributed random functions $\bm{Y}$. In addition to enabling causal identification, non-Gaussian-processes are robust against outlying curves \citep{zhu2011robust}. For finite sample inference, we truncate the orthonormal basis at level $K$ such that $\bm{\phi}_j = (\phi_{j1}, \ldots, \phi_{jK})^T$, as commonly done in existing functional data analysis literature. Consequently, \eqref{eq2} is turned into
\begin{align} \label{eq3}
W_{j}(m) = \sum_{k = 1}^K Z_{jk} \phi_{jk}(\omega_{j}(m)) + e_{j}(m).
\end{align}
 
\subsection{Causal Identifiability} \label{sec:ci}

The proposed functional BNs are useful representations of directed conditional independence and causal relationships for multivariate functional data. The big remaining question is the learning of the underlying (causal) DAGs from observational data. Constraint-based methods, which are often model-free, have been popular for DAG learning. For the proposed functional BNs, we, in principle, can also use constraint-based methods, which test for conditional independence of pairs of functions. However, conditional independence tests are notoriously difficult and inefficient even for scalar random variables. Furthermore, even if we have access to oracle conditional independence tests for random functions, we can only hope for identifying the best MEC by definition (recall that an MEC contains DAGs with exactly the same set of conditional independence relationships). This may be acceptable if one is only interested in learning conditional independence relationships. But as mentioned in Section \ref{sec::introduction}, for causal discovery, this is clearly unsatisfactory because the directionality of a potentially large number of edges of Markov equivalent DAGs may be left undetermined and hence the causal interpretations of these edges are unclear. Because the proposed FLiNG-BN is a proper probability model, we can exploit certain feature of the model, namely the non-Gaussianity, to uniquely identify the underlying causal DAG.
 
\begin{definition}[Causal Identifiability]
Suppose $\bm{Y}$ follows the FLiNG-BN $\mathcal{B} = (G, P)$, and suppose $\bm{W}$ is a noisy version of $\bm{Y}$ with noise variances $\bm{\sigma}= (\sigma_1, \ldots, \sigma_p)$ as defined in \eqref{eq2}. Let $P_W$ denote the distribution of $\bm{W}$ induced from FLiNG-BN and the noises. We say that the causal DAG of FLiNG-BN is identifiable from $\bm{W}$ if there does not exist another BN $\mathcal{B}' = (G', P')$ where $G'\neq G$ and noise variances $\bm{\sigma}' = (\sigma'_1, \ldots, \sigma'_p)$ such that the induced distributions on $\bm{W}$, $P'_W$, is equivalent to $P_W$, i.e., $P_W(\bm{W}) \equiv P'_W(\bm{W})$.
\end{definition}

\begin{theorem}[Causal Identifiability] \label{them1}
The causal DAG of FLiNG-BN is identifiable if the number of Gaussian mixture components $M_{jk} > 1, \forall j, k$.
\end{theorem}

Theorem \ref{them1} signifies that by examining the probability distribution $P_W$, to which we have access through the observational data alone, one can gauge the likelihood that a given causal DAG is the data generating DAG. With a finite dataset, we shall focus on weighing different candidate causal DAGs by their posterior probabilities. Here, we provide the outline of the proof; the complete proof is given in Section A of the Supplementary Material. Given a chosen set of basis functions, we show the result in the space of basis coefficients. The problem then transforms to prove that, given $\bm{Z} = \bm{B} \bm{Z} + \bm{\epsilon}$ and observe $\bm{W} = \bm{Z} + \bm{e}$, there does not exist another equivalent parameterization $\bm{Z}' = \bm{B}' \bm{Z}' + \bm{\epsilon}'$ and $\bm{W} = \bm{Z}' + \bm{e}'$. Since we assume each component of $\bm{\epsilon}$ follows a Gaussian scale mixture, the induced distribution on $\bm{W}$ is a multivariate Gaussian mixture (with different precision matrices). We then prove the causal effect matrix $\bm{B}$ is uniquely identifiable from such mixture model by combining the identification of Gaussian mixture components, uniqueness of LDL decomposition, and proof of causal ordering identification. We demonstrate the identifiability result with a toy example.

\begin{figure}[h]
\centering
\includegraphics[width=1\linewidth]{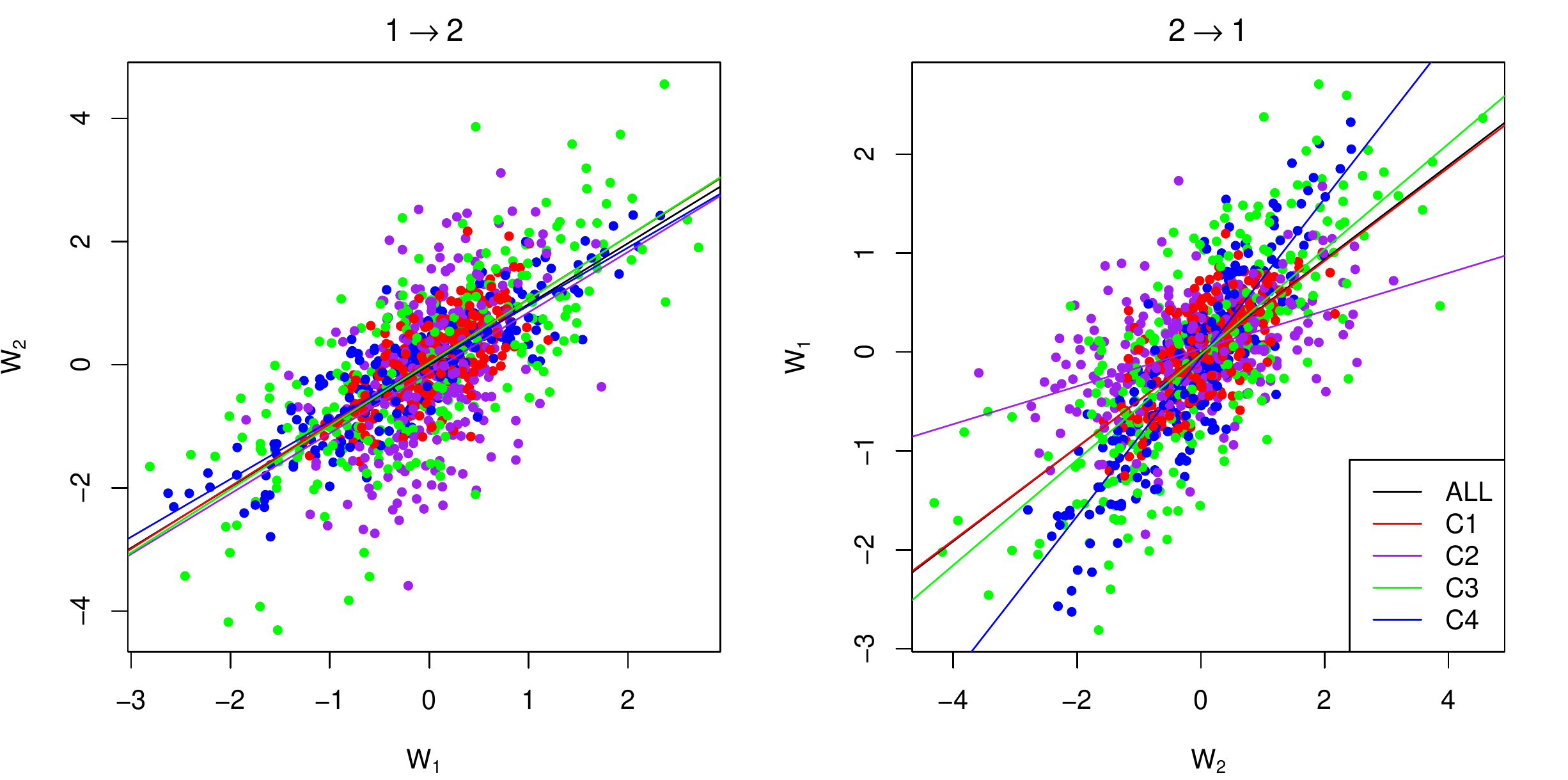}
\caption{A toy example for demonstration of causal identification. The left (right) panel shows the linear regression of $W_2$ ($W_1$) on $W_1$ ($W_2$). Data are simulated from the causal graph $1\to 2$. Colored lines are the fitted linear regressions for all observations and for the observations in groups $C_1$--$C_4$.}
\label{demo}
\end{figure}

\begin{example} \label{exp1}
Consider a true functional causal graph $1\to 2$ and the corresponding data generating model $Z_1 = \epsilon_1$ with $\epsilon_1 \sim 0.5 N(0, 0.5) + 0.5 N(0, 1)$ and $Z_2 = Z_1 + \epsilon_2$ with $\epsilon_2 \sim 0.5 N(0, 0.5) + 0.5 N(0, 1)$; note for simplicity, we assume in this example that the number of basis functions is $K=1$. Assume we observe with noises $W_1 = Z_1 + e_1$ and $W_2 = Z_2 + e_2$ with $e_1, e_2 \sim N(0, 0.1)$. We sample $n = 1000$ observations from this model and index them by the subscript $i=1,\dots,n$. For the purpose of illustration, suppose we know the mixture component assignment of each observation and define four groups of observations based on the combination of variances of $\epsilon_1$ and $\epsilon_2$,
\begin{equation*}
\begin{aligned}
& C_1 = \{i: \text{Var}(\epsilon_{i1}) = 0.5 ~\text{and}~ \text{Var}(\epsilon_{i2}) = 0.5\}, ~ C_2 = \{i: \text{Var}(\epsilon_{i1}) = 0.5 ~\text{and}~ \text{Var}(\epsilon_{i2}) = 1\}, \\
& C_3 = \{i: \text{Var}(\epsilon_{i1}) = 1 ~\text{and}~ \text{Var}(\epsilon_{i2}) = 0.5\}, ~ C_4 = \{i: \text{Var}(\epsilon_{i1}) = 1 ~\text{and}~ \text{Var}(\epsilon_{i2}) = 1\}.
\end{aligned}
\end{equation*}
We fit linear regression separately to all observations and to observations in each of the four groups with the true causal direction $1 \to 2$ (regressing $W_2$ on $W_1$) and the  anti-causal direction $2 \to 1$ (regressing $W_1$ on $W_2$), which are shown in Figure \ref{demo}. We observe that the fitted lines are almost identical in the causal direction for all groups whereas they can be quite different across groups in the anti-causal direction. Therefore, only the true causal graph gives a unique regression coefficient among all groups. Notice that if there is only one mixture component (i.e., degeneration to the Gaussian case), no comparison can be made between the causal and anti-causal directions since there will be only one regression line.
\end{example}

The next counter example illustrates the necessity of the non-Gaussian assumption for causal identification.

\begin{example}
Consider a similar bivariate case to Example \ref{exp1} but now the exogenous variables are Gaussian instead of mixture of Gaussian. Suppose the true functional causal graph $1\to 2$ and the corresponding data generating model $Z_1 = \epsilon_1$ with $\epsilon_1 \sim N(0, \tau_1)$ and $Z_2 = b Z_1 + \epsilon_2$ with $\epsilon_2 \sim N(0, \tau_2)$; note again for simplicity, we assume in this example that the number of basis functions is $K = 1$. Assume we observe with noises $W_1 = Z_1 + e_1$ and $W_2 = Z_2 + e_2$ with $e_1 \sim N(0, \sigma_1)$ and $e_2 \sim N(0, \sigma_2)$. The induced joint distribution on $\bm{W} = (W_1, W_2)$ is then bivariate Gaussian with mean $0$ and covariance matrix
$$\begin{pmatrix}
\tau_1 + \sigma_1 & b \tau_1\\
b \tau_1 & b^2 \tau_1 + \tau_2 + \sigma_2
\end{pmatrix}. $$

Further consider the anti-causal model $2 \to 1$ where $Z_2' = \epsilon_2'$ with $\epsilon_2' \sim N(0, \tau_2')$ and $Z_1' = b' Z_2' + \epsilon_1'$ with $\epsilon_1' \sim N(0, \tau_1')$. Suppose $W_1 = Z_1' + e_1'$ and $W_2 = Z_2' + e_2'$ with $e_1' \sim N(0, \sigma_1')$ and $e_2' \sim N(0, \sigma_2')$. The induced joint distribution on $\bm{W} = (W_1, W_2)$ is still bivariate Gaussian with mean 0 and covariance matrix
$$\begin{pmatrix}
b^{'2} \tau_2' + \tau_1' + \sigma_1' & b' \tau_2'\\
b' \tau_2' & \tau_2' + \sigma_2'
\end{pmatrix}. $$ 
For any chosen $\tau_1', \sigma_1' > 0$ such that $\tau_1' + \sigma_1' < \tau_1 + \sigma_1 - b^2 \tau_1^2 / (b^2 \tau_1 + \tau_2 + \sigma_2)$, if we set
\begin{equation*}
\begin{aligned}
& b' = (\tau_1 + \sigma_1 - \tau_1' - \sigma_1') / b \tau_1, \\
&\tau_2' = b^2 \tau_1^2 / (\tau_1 + \sigma_1 - \tau_1' - \sigma_1'), \\
& \sigma_2' = b^2 \tau_1 + \tau_2 + \sigma_2 - b^2 \tau_1^2 / (\tau_1 + \sigma_1 - \tau_1' - \sigma_1'),
\end{aligned}
\end{equation*}
then the induced distribution coincides with that under the true causal model (i.e., Gaussian with mean 0 and the same covariance). Therefore, causal identification fails in this case.
\end{example}

\section{Bayesian Inference} \label{sec::inference}

The inference of the proposed FLiNG-BN framework can be carried out in either a frequentist (e.g., maximizing penalized likelihood) or a Bayesian (e.g., sampling from posterior distribution) fashion. Existing frequentist functional graphical models \citep{qiao2019functional,qiao2020doubly,zapata2022partial,solea2022copula,lee2022nonparametric,lee2022functional} often estimate graphs in two separate steps -- (i) estimate the basis coefficient sequence of each function marginally via functional principle component analysis, and (ii) learn an directed/undirected graph based on the estimated coefficient sequences. However, the eigenfunctions that marginally explain the most variation of each individual function do not necessarily explain well the conditional/causal relationships among a set of functions. Moreover, the estimation uncertainty is not propagated from the first step to the second, which may result in overly confident inference. To mitigate these potential drawbacks of the two-step approaches, we propose a fully Bayesian inference procedure that jointly infers basis coefficient sequences and the DAG structure. This joint inference approach constructs orthonormal basis functions adaptive to their conditional/causal relationships and allows for finite-sample inference and uncertainty quantification. 

\subsection{Adaptive Orthonormal Basis Functions}

We assume the basis functions to be shared across all random functions \citep{kowal2017bayesian,zapata2022partial}, $\phi_{jk}(\omega) := \phi_k(\omega),\forall j\in [p]$, which are more parsimonious than models based on function-specific basis functions. Moreover, the common basis functions put the basis coefficient sequences $\bm{Z}_j,\forall j\in[p]$ on an equal footing (e.g., the magnitudes of basis coefficients are directly comparable) so that they are directly comparable and the BN on $\bm{Z}$ has a more coherent interpretation. In this case, the non-zero matrix block $\bm{B}_{j\ell}$ corresponds to a causal connection from $\bm{Y}_\ell$ to $\bm{Y}_j$. Loosely speaking, if we regard the basis functions as signal channels, then a significant non-zero $B_{j\ell}(k_j, k_\ell)$ indicates that $\bm{Y}_\ell$ directly affects $\bm{Y}_j$ through its signal transmission from the $k_\ell$-th channel to the $k_j$-th channel. 

As mentioned above, we do not pre-specify a fixed set of orthonormal basis functions but instead they are learned adaptively from data by further expanding them with spline basis functions \citep{kowal2017bayesian}, $\phi_k(\omega) = \sum_{\ell = 1}^L A_{k\ell} b_\ell(\omega)$, where $\bm{b} = (b_1, \ldots, b_L)^T$ is a set of cubic B-spline basis functions with equally spaced knots and $\bm{A}_k = (A_{k1}, \ldots, A_{kL})^T, \forall k \in [K]$ are spline coefficients. Because $\bm{A}_k$'s are not fixed \emph{a priori}, so are $\phi_k$'s.

\subsection{Prior Model}

\begin{figure}[h]
\centering
\includegraphics[width=1\linewidth]{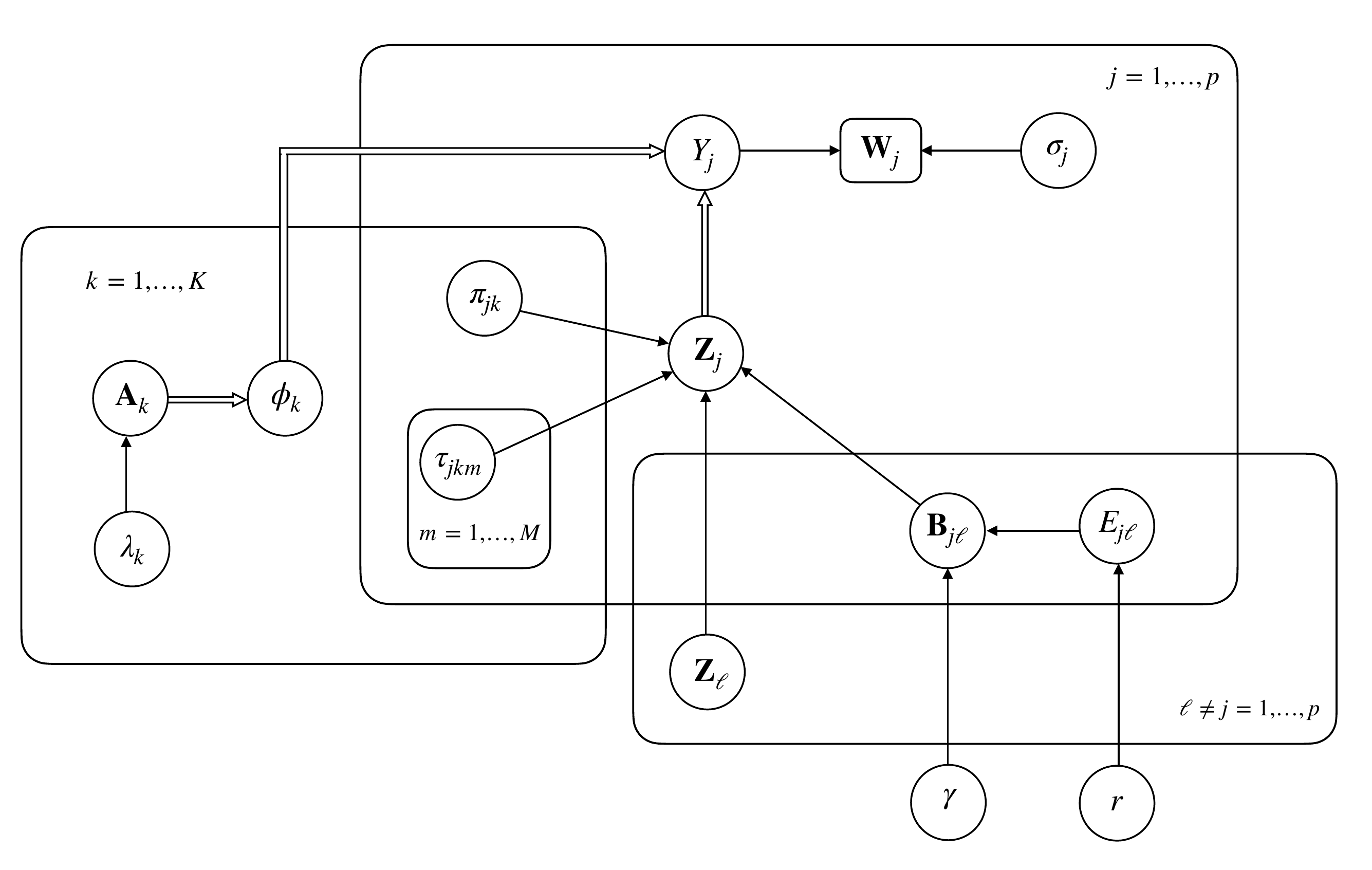}
\caption{A DAG illustrating the model hierarchy. Single-line arrows are stochastic relationships and double-line arrows are deterministic relationships. The observed node $\bm{W}_j$ is shown in rectangle and other nodes are shown in circles.}
\label{hmodel}
\end{figure}

We summarize our model and its entailed parameters using a DAG shown in Figure \ref{hmodel}. The prior distributions of the model parameters are introduced in this section. We simulate posterior samples through Markov chain Monte Carlo (MCMC). Details are given in Section B of the Supplementary Material. 

\paragraph{Prior on B-spline Coefficients $\bm{A}_k$} The prior on $\bm{A}_k$ serves three purposes. First, it forces $\phi_k$'s to be orthonormal, i.e., $\int \phi_k(\omega) \phi_h(\omega) d\omega = I(k = h), \forall k, h \in[K]$. Second, it regularizes the roughness of $\phi_k$'s to prevent overfitting and sorts the orthonormal basis functions by increasing roughness. Third, it enables posterior inference on the orthonormal basis functions simultaneously with the graph estimation without having to fix them \emph{a priori}.

We summarize the main steps of prior specification and refer the details to \cite{kowal2017bayesian}. First, to regularize the roughness of $\phi_k$ in a frequentist framework, one would consider a penalized likelihood with the roughness penalty,
$$\lambda_k \mathcal{P}(\bm{A}_k) = \lambda_k \int [\phi_k^{''}(\omega)]^2 \ d\omega = \lambda_k \bm{A}_k^T \bm{\Omega} \bm{A}_k,$$
where $\lambda_k > 0$ is the regularization parameter and $\bm{\Omega} = \int \bm{b}^{''}(\omega)[\bm{b}^{''}(\omega)]^T \ d\omega$. As a Bayesian counterpart, the regularization term is equivalent to a prior on the B-spline coefficients $\bm{A}_k \sim N(\bm{0}, \lambda_k^{-1} \bm{\Omega}^{-})$, where $\bm{\Omega}^{-}$ is a pseudo-inverse (since $\bm{\Omega}$ is rank-deficient by 2). Let $\bm{\Omega} = \bm{U} \bm{D} \bm{U}^T$ be the singular value decomposition of $\bm{\Omega}$. To facilitate efficient computation, we follow \cite{wand2008semiparametric} and reparameterize $\phi_k = \sum_{\ell=1}^L A_{k\ell}b_\ell = \sum_{\ell=1}^L \tilde{A}_{k\ell} \tilde{b}_\ell$ with $\tilde{\bm{b}}(\omega) = (1, \omega, \bm{b}^T(\omega) \bm{U}_P \bm{D}_P^{-1/2})^T$ where $\bm{D}_P$ is the $(L - 2) \times (L - 2)$ submatrix of $\bm{D}$ corresponding to non-zero singular values and $\bm{U}_P$ is the corresponding $L \times (L - 2)$ submatrix of $\bm{U}$. The reparameterization induces a prior on $\tilde{\bm{A}}_k = (\tilde{A}_{k1}, \ldots, \tilde{A}_{kL})^T\sim N(0, S_k)$ where $S_k = \mathrm{diag}(\infty, \infty, \lambda_k^{-1}, \ldots, \lambda_k^{-1})$ with the first two dimensions corresponding to the unpenalized constant and linear terms. In practice, one can replace $\infty$ by a large number, say $10^8$. 

Second, we constrain the regularization parameters $\lambda_1 > \cdots > \lambda_K > 0$ to identify the ordering of basis functions, which sorts the basis functions by decreasing smoothness. Unlike the functional principal component analysis (PCA) where the principal components are ordered by the proportion of variance explained, the adopted Bayesian approach is less prone to rough functions. Given the ordering constraint, a uniform prior is imposed such that $\lambda_k \sim U(L_k, U_k)$, where $U_1 = 10^8$, $L_k = \lambda_{k + 1}$ for $k = 1, \ldots, K - 1$, $U_k = \lambda_{k - 1}$ for $k = 2, \ldots, K$, and $L_K = 10^{-8}$.

Finally, consider the orthonormal constraint
\begin{align} \label{penmt}
\int \phi_k(\omega) \phi_h(\omega) = \int \tilde{\bm{A}}_k^T \tilde{\bm{b}}(\omega) \tilde{\bm{b}}^T(\omega)\tilde{\bm{A}}_h \ d\omega = \tilde{\bm{A}}_k^T \bm{J} \tilde{\bm{A}}_h = I(k = h),
\end{align}
with $\bm{J} = \int \tilde{\bm{b}}(\omega) \tilde{\bm{b}}^T(\omega) \ d\omega$. This constraint can be easily enforced by projection and normalization during the course of MCMC; see Section B of the Supplementary Material for details.

\paragraph{Priors on the DAG Adjacency Matrix $\bm{E}$ and Direct Causal Effects $\bm{B}$} The key problem we aim to address in this article is causal structure learning, i.e., inferring the adjacency matrix, $\bm{E} = (E_{j\ell})$ (recall $E_{j \ell} = 1$ if and only if $\ell \to j$). We propose to use a beta-Bernoulli-like prior $E_{j \ell} \sim \mathrm{Bernoulli}(r)$ with $r \sim \mathrm{Beta}(a_r, b_r)$, subject to the acyclicity constraint,
$$P(\bm{E} | r) \propto \prod_{j\neq\ell} r^{E_{j\ell}} (1-r)^{1-E_{j\ell}} I(G \text{ is a DAG}).$$
We set $a_r = b_r = 1$. \cite{scott2010bayes} showed that the beta-Bernoulli prior allows automatic multiplicity adjustment in sparse regression problem. In our context, the marginal distribution of $\bm{E}$ with $r$ integrated out equals
\begin{align} \label{mar}
P(\bm{E}) \propto \mathrm{Beta}\left(\sum_{j\neq\ell}E_{j\ell}+1,\sum_{j\neq\ell}(1-E_{j\ell})+1\right)I(G \text{ is a DAG}).
\end{align}
The marginal distribution strongly prevents false discoveries by increasing the penalty against additional edges as the dimension $p$ grows. For example, the marginal \eqref{mar} favors an empty graph over a graph with one edge by a factor of $p^2 - p$, which increases with $p$.

Conditional on $\bm{E}$, we assume independent matrix-variate spike-and-slab priors on the direct causal effects,
\begin{align*}
\bm{B}_{j \ell}|E_{j \ell} \sim (1 - E_{j \ell}) \delta_{\bm{O}}(\bm{B}_{j \ell}) + E_{j \ell} N(\bm{B}_{j\ell}|\bm{O}, \gamma \bm{I}, \bm{I}),
\end{align*}
where $\delta_{\bm{O}}(\cdot)$ is a point mass at a $K\times K$ zero matrix $\bm{O}$ and $N(\cdot|\bm{O}, \gamma \bm{I}, \bm{I})$ is a centered matrix-variate normal distribution with row and column covariance matrices $\gamma \bm{I}$ and $\bm{I}$ where $\bm{I}$ is a $K\times K$ identity matrix. The hyperparameter $\gamma$ indicates the overall causal effect size and is assumed to follow a conjugate inverse-gamma prior, $\gamma \sim IG(a_\gamma, b_\gamma)$ with $a_\gamma = b_\gamma = 1$.

\paragraph{Prior on the Gaussian Scale Mixture} We choose conjugate priors,
\begin{align*}
\bm{\pi}_{jk} = (\pi_{jk1}, \ldots, \pi_{jkM}) \sim \text{Dirichlet}(\alpha,\dots,\alpha), ~ \tau_{jkm} \sim IG(a_\tau, b_\tau), ~ \forall j \in [p], k \in [K], m \in [M],
\end{align*}
which allows for straightforward Gibbs sampling. As default, we set $\alpha = 1$ and $a_\tau = b_\tau = 1$. 

\paragraph{Prior on Observation Noises} We complete the prior specification with a conjugate inverse-gamma prior on the variance of observation noises, $\sigma_j \sim IG(a_\sigma, b_\sigma), \forall j \in [p]$ with $a_\sigma = b_\sigma = 0.01$.

Finally, we summarize the differences between the proposed FLiNG-BN and the work from \cite{lee2022functional}. First, \cite{lee2022functional} assume  their functions to be noiseless whereas we consider the scenario where functions are observed with noises. The causal identifiability theory is significantly more complicated when functions are noisy. Second, they assume their functions to be Gaussian whereas our functions are non-Gaussian; this difference leads to different learning algorithms and identifiability theory. Third, their inference is a two-step procedure based on causal ordering identification and sparse function-on-function regression, while the proposed Bayesian hierarchical model admits one-step inference procedure, which learns the graph structure by directly searching in the graph space without having to learn the causal ordering first.

\section{Simulation Studies} \label{sec::experiment}

We conducted simulation studies to evaluate the proposed FLiNG-BN model. We considered two scenarios. In the first scenario, the functions were observed on an evenly spaced grid; this is the scenario commonly studied in the existing functional undirected graphical models \citep{qiao2019functional} and is also similar to our later EEG application. In the second scenario, the functions were observed on an unevenly spaced grid, similar to the COVID-19 longitudinal application (the details are shown in Section C of the Supplementary Material). We compared the proposed FLiNG-BN with a functional undirected graphical model (FGLASSO; \citealt{qiao2019functional}). We did not make comparison with \cite{lee2022functional} due to lack of publicly available code at the time of submission. In addition, we compared FLiNG-BN with approaches based on two-step estimation procedures. In the first step, we extracted basis coefficients obtained from functional PCA using the package \texttt{fdapace} \citep{carroll2021}. In the second step, given the estimated basis coefficients, we constructed causal graphs using either the LiNGAM \citep{shimizu2006linear} algorithm (termed FPCA-LiNGAM) or the PC \citep{spirtes1991algorithm} algorithm (termed FPCA-PC). LiNGAM estimates a causal DAG based on the linear non-Gaussian assumption whereas PC generally returns only an equivalence class of DAGs based on conditional independence tests. Their implementations are available from R package \texttt{pcalg} \citep{kalisch2020overview}.

To mimic the EEG data application, we simulated data from FLiNG-BN with all the combinations of sample size $n \in \{50, 100, 200\}$, number of functions $p \in \{30, 60, 90\}$, and grid size $d \in \{125, 250\}$. The grid spanned the unit interval $[0,1]$. We set the true number of basis functions to be $K = 5$. To generate basis functions, we first simulated the non-orthnormal functions $\phi_k^U, \forall k \in [K]$ from a set of $L = 6$ cubic B-spline basis functions with evenly spaced knots, $\phi_k^U = \sum_{\ell=1}^L A_{k\ell}b_\ell$, where $A_{k\ell}$'s were generated from a standard normal distribution. We then empirically orthonormalized $(\phi_1^U, \ldots, \phi_K^U)$ to get the orthonormal basis functions $(\phi_1, \ldots, \phi_K)$. The simulation true causal graph $G$ was generated from the Erd\H{o}s-R\'{e}nyi model with connection probability $2/p$, subject to the acyclicity constraint. Given the true graph $G$, each block of non-zero direct causal effects $\bm{B}_{j\ell}$ was generated independently from a standard matrix-variate normal distribution. Then the basis coefficient sequences $\bm{Z}$ were generated from \eqref{eq1} where the exogenous variables $\bm{\epsilon}_j$'s were generated from a centered Laplace distribution with scale $b = 0.5$. Note that when we fit FLiNG-BN to the simulated data, we still assumed the exogenous variables to be discrete scale Gaussian mixture although the simulation true exogenous variables were Laplace (i.e., continuous scale Gaussian mixture). Finally, noisy observations were simulated following \eqref{eq3} with the signal-to-noise ratio, i.e., the mean value of $|y_j^{(i)}(\omega_j^{(i)}(m))| / \sigma_j$ across all samples $i \in [n]$ and grid points $m \in [m_{j}^{(i)}]$, set to 5.

For implementing the proposed FLiNG-BN, we set the number of mixture components to $M = 5$ and the number B-spline basis functions to $L=20$ (note that the simulation truth was $L=6$), and ran MCMC for 5,000 iterations (discarding the first half as burn-in and retaining every 5th iteration after burn-in). The causal graph $G$ was estimated by thresholding the posterior probability of inclusion at 0.5 (i.e., the median probability model). Parameters of competing methods were set to their default values. To assess the graph recovery performance, we calculated true positive rate (TPR), false discovery rate (FDR), and Matthews correlation coefficient (MCC),
\begin{align*}
& \text{TPR} = \text{TP}(\text{TP} + \text{FN})^{-1}, ~~~~~ 
\text{FDR} = \text{FP}(\text{TP} + \text{FP})^{-1}, \\
& \text{MCC} = (\text{TP} \times \text{TN} - \text{FP} \times \text{FN})\left[(\text{TP} + \text{FP}) \times (\text{TP} + \text{FN}) \times (\text{TN} + \text{FP}) \times (\text{TN} + \text{FN})\right]^{-1/2},
\end{align*}
where TP, TN, FP, and FN stand for the numbers of true positives, true negatives, false positives, and false negatives, respectively. MCC ranges from $-$1 to 1 with $0$ indicating a random guess and $1$ a perfect recovery. Since FGLASSO learns an undirected graph, we compared it with a moralization of the true graph\footnote{Graph moralization converts a DAG to an undirected graph by first marrying all the unmarried parents and then removing all the directions. A probability distribution that respects the Markov property of a DAG must respect the Markov property of its moral graph.}. Similarly, since PC algorithm returns the MEC representation\footnote{The MEC representation is shown in an essential graph, where any edge presented between two nodes is directed if and only if it follows the same direction in all members of this MEC. Otherwise, it is undirected.}, we compared it with the MEC of the true causal graph. 

The results based on 50 repeat simulations are summarized in Table \ref{s1}, from which we conclude that the proposed FLiNG-BN significantly outperformed all the competitors FGLASSO, FPCA-LiNGAM, and FPCA-PC across all combinations of $n$, $p$, and $d$. This is not surprising because (i) FGLASSO is not designed for learning directed graphs; they were compared with the proposed FLiNG-BN because of lack of alternative functional BN implementation. (ii) Although FPCA-LiNGAM and FPCA-PC are capable of learning directed graphs, they still performed poorly because they are implemented in a two-step procedure where there is little reason to believe that the basis coefficients extracted by the functional PCA in the first step are useful to capture the functional dependence in the second step. (iii) Unlike the proposed approach, none of the competing methods controls for false discovery and some impose the stringent Gaussian assumption, resulting in high FDR and/or low TPR. Our suggested method to determine $K$ also worked well.

\begin{table}[ht]
\caption{Functions observed on evenly spaced grid. Average operating characteristics based on 50 repetitions are reported; standard deviations are given within the parentheses. Since LiNGAM is not applicable to cases where $q > n$ with $q = pK$ being the total number of extracted basis coefficients across all functions, the results from those cases are not available and indicated by -.}
\resizebox{\textwidth}{!}{
\centering
\begin{tabular}{ccc|ccc|ccc|ccc|ccc}
\toprule
\multirow{2}{*}{$p$} & \multirow{2}{*}{$d$} & \multirow{2}{*}{$n$} & \multicolumn{3}{c|}{FLiNG-BN} & \multicolumn{3}{c|}{FGLASSO} & \multicolumn{3}{c|}{FPCA-LiNGAM} & \multicolumn{3}{c}{FPCA-PC} \\
\cmidrule(lr){4-6} \cmidrule(lr){7-9} \cmidrule(lr){10-12} \cmidrule(lr){13-15}
& & & TPR & FDR & MCC & TPR & FDR & MCC & TPR & FDR & MCC & TPR & FDR & MCC \\
\midrule
30 & 125 & 50 & 0.62 (0.07) & 0.14 (0.07) & 0.72 (0.07) & 0.58 (0.02) & 0.88 (0.02) & 0.16 (0.02) & - & - & - & 0.22 (0.03) & 0.89 (0.02) & 0.12 (0.02) \\
30 & 125 & 100 & 0.71 (0.08) & 0.19 (0.05) & 0.75 (0.06) & 0.63 (0.03) & 0.85 (0.03) & 0.20 (0.03) & 0.84 (0.02) & 0.85 (0.01) & 0.31 (0.02) & 0.30 (0.01) & 0.89 (0.01) & 0.13 (0.01) \\
30 & 125 & 200 & 0.73 (0.05) & 0.13 (0.08) & 0.79 (0.06) & 0.69 (0.03) & 0.84 (0.05) & 0.19 (0.03) & 0.92 (0.04) & 0.87 (0.01) & 0.30 (0.01) & 0.13 (0.02) & 0.96 (0.01) & 0.02 (0.01) \\
\midrule
30 & 250 & 50 & 0.68 (0.05) & 0.25 (0.08) & 0.73 (0.06) & 0.57 (0.02) & 0.88 (0.04) & 0.16 (0.04) & - & - & - & 0.30 (0.02) & 0.87 (0.01) & 0.15 (0.01) \\
30 & 250 & 100 & 0.75 (0.04) & 0.26 (0.03) & 0.74 (0.03) & 0.64 (0.03) & 0.85 (0.04) & 0.18 (0.03) & 0.88 (0.04) & 0.86 (0.02) & 0.34 (0.02) & 0.18 (0.02) &0.92 (0.02) & 0.08 (0.01) \\
30 & 250 & 200 & 0.85 (0.01) & 0.30 (0.06) & 0.79 (0.04) & 0.69 (0.02) & 0.83 (0.04) & 0.21 (0.02) & 0.97 (0.05) & 0.85 (0.03) & 0.35 (0.03) & 0.22 (0.02) & 0.94 (0.01) & 0.08 (0.02) \\
\midrule
60 & 125 & 50 & 0.68 (0.03) & 0.05 (0.03) & 0.80 (0.02) & 0.57 (0.04) & 0.89 (0.06) & 0.11 (0.05) & - & - & - & 0.28 (0.02) & 0.87 (0.01) & 0.16 (0.01) \\
60 & 125 & 100 & 0.68 (0.04) & 0.12 (0.04) & 0.75 (0.04) & 0.60 (0.03) & 0.85 (0.05) & 0.15 (0.04) & - & - & - & 0.28 (0.01) & 0.89 (0.01) & 0.15 (0.01) \\
60 & 125 & 200 & 0.74 (0.02) & 0.11 (0.02) & 0.82 (0.02) & 0.61 (0.03) & 0.82 (0.04) & 0.17 (0.03) & 0.86 (0.03) & 0.89 (0.02) & 0.25 (0.02) & 0.22 (0.01) & 0.95 (0.01) & 0.11 (0.01) \\
\midrule
60 & 250 & 50 & 0.70 (0.02) & 0.15 (0.02) & 0.77 (0.01) & 0.59 (0.04) & 0.82 (0.04) & 0.16 (0.03) & - & - & - & 0.35 (0.02) & 0.85 (0.02) & 0.21 (0.01) \\
60 & 250 & 100 & 0.70 (0.01) & 0.13 (0.10) & 0.79 (0.05) & 0.62 (0.04) & 0.80 (0.04) & 0.17 (0.03) & - & - & - & 0.26 (0.01) & 0.89 (0.02) & 0.13 (0.01) \\
60 & 250 & 200 & 0.76 (0.02) & 0.11 (0.01) & 0.85 (0.01) & 0.69 (0.05) & 0.80 (0.03) & 0.19 (0.04) & 0.91 (0.02) & 0.85 (0.02) & 0.33 (0.01) & 0.17 (0.01) & 0.84 (0.05) & 0.15 (0.03) \\
\midrule
90 & 125 & 50 & 0.63 (0.04) & 0.10 (0.04) & 0.75 (0.03) & 0.52 (0.02) & 0.89 (0.03) & 0.10 (0.03) & - & - & - & 0.23 (0.01) & 0.88 (0.00) & 0.15 (0.01) \\
90 & 125 & 100 & 0.66 (0.03) & 0.12 (0.03) & 0.74 (0.02) & 0.55 (0.04) & 0.87 (0.03) & 0.15 (0.02) & - & - & - & 0.18 (0.01) & 0.92 (0.01) & 0.10 (0.01) \\
90 & 125 & 200 & 0.67 (0.02) & 0.13 (0.02) & 0.76 (0.01) & 0.57 (0.03) & 0.85 (0.04) & 0.17 (0.03) & - & - & - & 0.17 (0.01) & 0.94 (0.01) & 0.08 (0.01) \\
\midrule
90 & 250 & 50 & 0.58 (0.03) & 0.09 (0.02) & 0.68 (0.03) & 0.54 (0.05) & 0.87 (0.04) & 0.11 (0.04) & - & - & - & 0.32 (0.01) & 0.84 (0.00) & 0.21 (0.01) \\
90 & 250 & 100 & 0.65 (0.05) & 0.13 (0.04) & 0.73 (0.04) & 0.58 (0.04) & 0.82 (0.03) & 0.15 (0.03) & - & - & - & 0.18 (0.02) & 0.93 (0.01) & 0.11 (0.01) \\
90 & 250 & 200 & 0.70 (0.02) & 0.12 (0.02) & 0.78 (0.01) & 0.61 (0.06) & 0.80 (0.04) & 0.18 (0.05) & - & - & - & 0.22 (0.01) & 0.89 (0.01) & 0.16 (0.01) \\
\bottomrule					
\end{tabular}}
\label{s1}
\end{table}

The proposed FLiNG-BN has a few hyperparameters $L$, $M$, $\alpha$, $(a_r, b_r)$, $(a_\gamma, b_\gamma)$, $(a_\tau, b_\tau)$, and $(a_\sigma, b_\sigma)$. We performed sensitivity analyses of these parameters at four different values with $(n, p, d) = (100, 30, 250)$. Results are summarized in Section C of the Supplementary Material. Our model appeared to be relatively robust within the tested ranges of hyperparameters.

\section{Applications} \label{sec::eeg}

We applied the proposed FLiNG-BN to the brain EEG dataset downloaded from \url{https://archive.ics.uci.edu/ml/datasets/eeg+database} \citep{zhang1995event}. The dataset consists of 122 subjects with 77 in the alcoholic group and 45 in the control group, and was previously used to demonstrate functional undirected graphical models by \cite{zhu2016bayesian} and \cite{qiao2019functional}. The 64 electrodes placed on subjects' scalps (standard positions) measuring voltage values were sampled at 256 Hz for one second. Each subject completed 120 trials under one stimulus or two stimuli. See \cite{zhang1995event} for details of the data collection procedure. We averaged all trials for each subject under the one stimulus condition. We separately analyzed these two groups to find their commonalities and differences of brain activity. Hence, we had $n = 77$ or $n = 45$ subjects and $p = 64$ functions representing the brain EEG signals at different scalp positions recorded at $d = 256$ time points. We focused on EEG signals filtered at $\alpha$ frequency bands between 8 and 12.5 Hz using the \texttt{eegfilt} function in the EEGLAB toolbox from MATLAB \citep{delorme2004eeglab}.

To check the Gaussianity of the observed functions, we performed Shapiro--Wilk normality test \citep{shapiro1965analysis} to each of $p = 64$ scalp positions at each of $d = 256$ time points. The null hypothesis (i.e., the observations are marginally Gaussian) was rejected for many combinations of scalp position and time point and therefore, the non-Gaussianity of the proposed model is deemed appropriate.

\begin{figure}[ht]
\centering
\begin{subfigure}[ht]{1 \textwidth}
\centering
\includegraphics[width = 1 \textwidth]{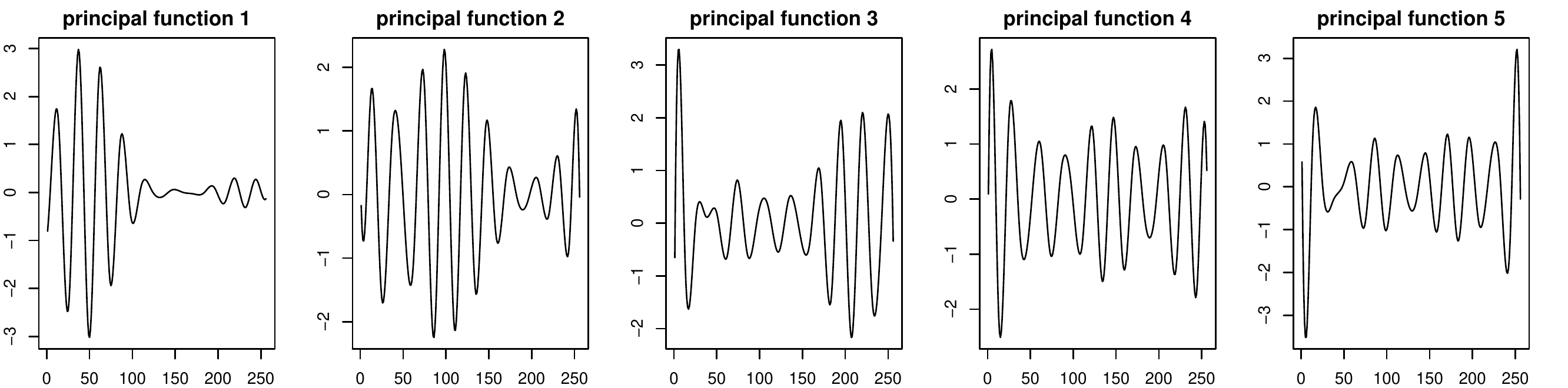}
\caption{Alcoholic group.}
\end{subfigure}
\begin{subfigure}[ht]{1 \textwidth}
\centering
\includegraphics[width = 1 \textwidth]{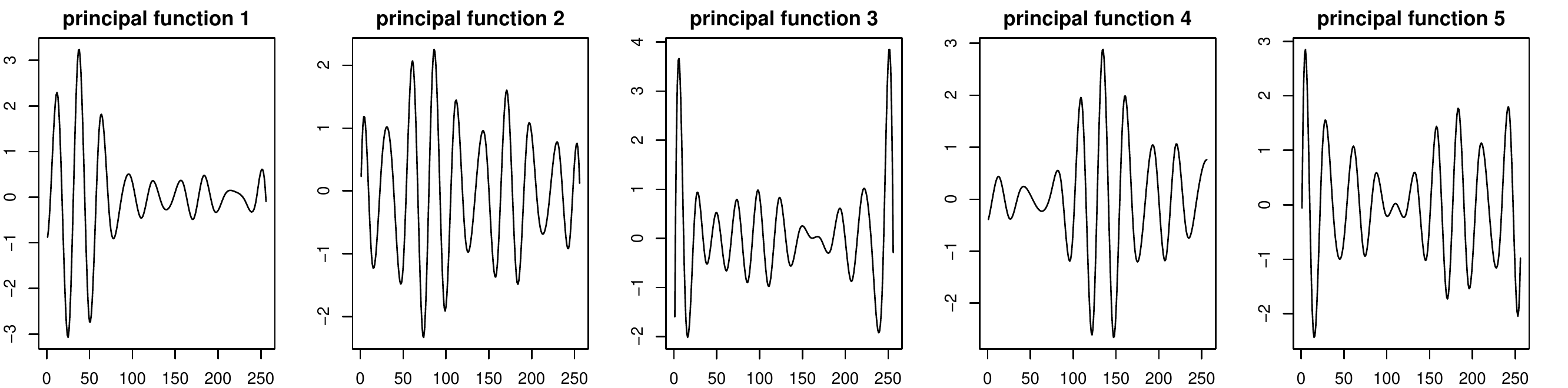}
\caption{Control group.}
\end{subfigure}
\caption{Estimated basis functions from brain EEG records that explained 90\% of the variation.}
\label{pfunction}
\end{figure}

\begin{figure}[ht]
\centering
\includegraphics[width = 1 \textwidth]{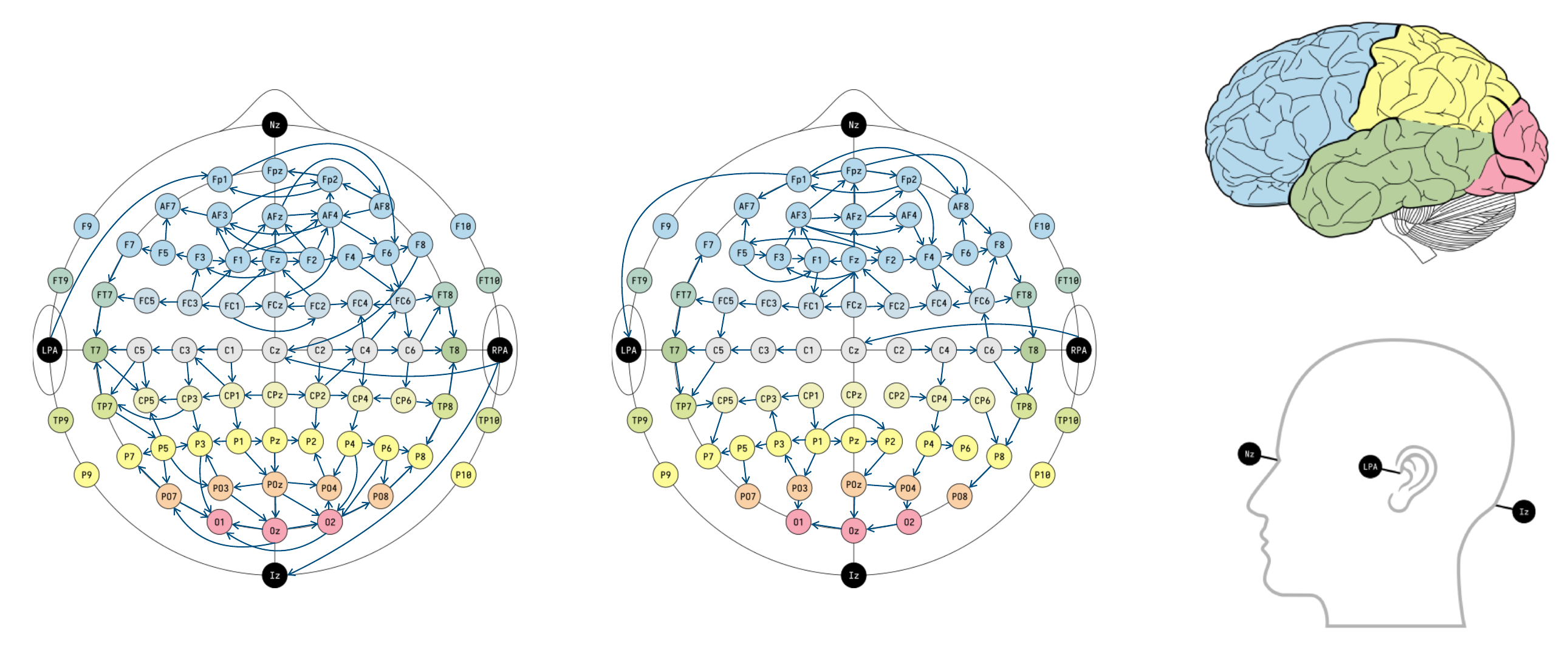}
\caption{Estimated causal brain networks from EEG records by FLiNG-BN with posterior probability of inclusion $\geq 0.9$, separately for the alcoholic (left) and control (right) group.}
\label{fnetwork}
\end{figure}

Five orthonormal basis functions were selected for both the alcoholic and control group according to the procedure described in Section B of the Supplementary Material. We ran MCMC for 10,000 iterations, discarded the first half as burn-in, and retained every 10th iteration after burn-in. The estimated basis functions are shown in Figure \ref{pfunction}. As evident from the plot, they are very similar across the two groups. The causal networks estimated by thresholding the posterior probability of inclusion at 0.9 are shown in Figure \ref{fnetwork}. The sparsity level is approximately 3.0\% for the alcoholic group and 2.5\% for the
control group. 

\begin{figure}[ht]
\centering
\includegraphics[width = 1 \textwidth]{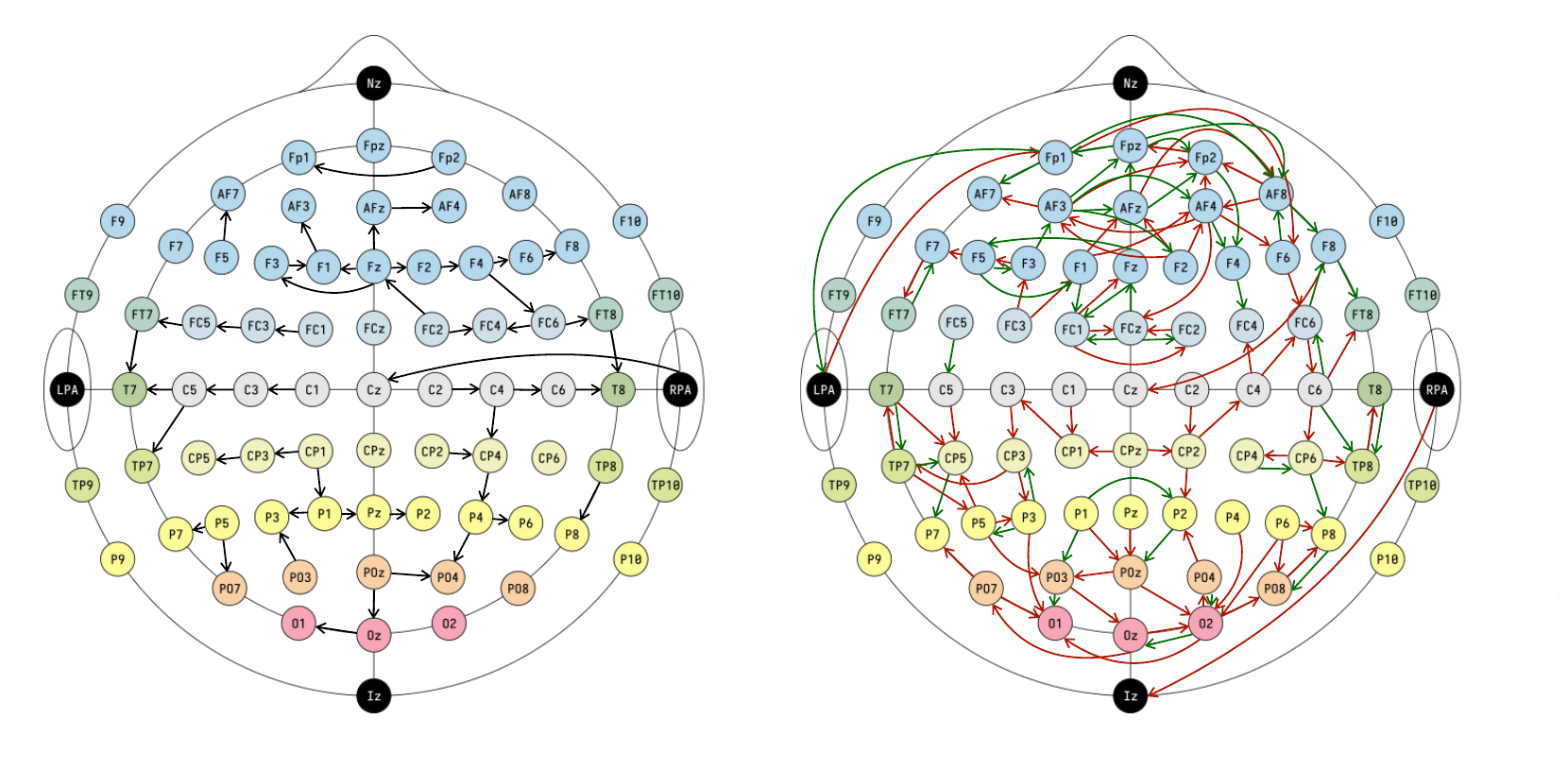}
\caption{Common (left panel) and differential (right panel) connections for the two groups. Black arrows indicate common connections, red arrows indicate connections detected by the alcoholic group only, and green arrows indicate connections detected by the control group only. }
\label{dnetwork}
\end{figure}

Our results reveal several interesting patterns. First, the connection is relatively dense in the frontal region for both groups. Second, the alcoholic group has more directed connections detected in the left temporal and occipital regions. Third, most brain locations tend to connect to adjacent positions, while distant locations are much less connected. Figure \ref{dnetwork} shows the common and differential networks for the two groups, where a substantial connectivity difference is observed between the two groups.

In addition, we demonstrated the proposed FLiNG-BN model with an additional application to COVID-19 multivariate longitudinal data, which have unevenly spaced measurements, in Section D of the Supplementary Material.

\section{Discussion} \label{sec::discussion}

In this paper, we have proposed a functional Bayesian network model for causal discovery from multivariate functional data. We have discussed in detail a specific case of functional Bayesian network, namely the functional linear non-Gaussian model, and proved the underlying causal structure is identifiable even if the functions are purely observational and observed with noises. A fully Bayesian inference procedure has been proposed to implement our framework. Through simulation studies and real data applications, we have demonstrated the ability of our model in causal discovery.

We briefly discuss several possible directions to extend our current work. First, we may replace the underlying DAG with cyclic graphs, chain graphs, or ancestral graphs for more general causal and conditional independence structures. We have chosen a linear non-Gaussian SEM on the basis coefficients but this model can be replaced with a nonlinear SEM. Second, instead of fixing the number of basis functions, one could resort to increasing shrinkage priors \citep{bhattacharya2011sparse, legramanti2020bayesian} to adaptively truncate redundant components. Finally, since we have two groups of observations in the EEG application, it would be interesting to jointly estimate the brain networks or directly estimate the differential network.

\bibliographystyle{jasa} 
\small \bibliography{reference.bib}

\end{document}